\documentstyle[aps,prl,preprint,floats,epsfig]{revtex}

\begin{document}

\preprint{\tighten\vbox{\hbox{\hfil CLNS 00/1661 }
                        \hbox{\hfil CLEO 00-01}
}}

\title{Search for {\boldmath $CP$} violation  in {\boldmath $B^{\pm} \to J/\psi \, K^{\pm}$} and  {\boldmath $B^{\pm} \to \psi(2S) \, K^{\pm}$}  decays}

\author{CLEO Collaboration}
\date{\today}

\maketitle
\tighten

\begin{abstract} 
We  present a search for direct  $CP$ violation in $B^{\pm} \to J/\psi
\, K^{\pm}$ and $B^{\pm} \to \psi(2S) \, K^{\pm}$ decays.  In a sample
of $9.7\times10^6$ $B \overline B$ meson pairs collected with the CLEO
detector,  we  have fully  reconstructed  534 $B^{\pm}   \to J/\psi \,
K^{\pm}$  and 120 $B^{\pm} \to \psi(2S)  \, K^{\pm}$  decays with very
low background. We have measured  the $CP$-violating charge  asymmetry
to be  $(+1.8\pm4.3[ {\rm stat}]\pm   0.4[{\rm syst}])\%$ for $B^{\pm}
\to J/\psi \, K^{\pm}$   and  $(+2.0\pm9.1[ {\rm  stat}]\pm   1.0[{\rm
syst}])\%$ for $B^{\pm} \to \psi(2S) \, K^{\pm}$. 
\end{abstract}
\newpage

{
\renewcommand{\thefootnote}{\fnsymbol{footnote}}

\begin{center}
G.~Bonvicini,$^{1}$ D.~Cinabro,$^{1}$ S.~McGee,$^{1}$
L.~P.~Perera,$^{1}$ G.~J.~Zhou,$^{1}$
E.~Lipeles,$^{2}$ S.~Pappas,$^{2}$ M.~Schmidtler,$^{2}$
A.~Shapiro,$^{2}$ W.~M.~Sun,$^{2}$
% ...
\end{center}
 
\small
\begin{center}
$^{1}${Wayne State University, Detroit, Michigan 48202}\\
$^{2}${California Institute of Technology, Pasadena, California 91125}\\
% ...
\end{center}
\setcounter{footnote}{0}
}
\newpage

\begin{figure}[htbp]
\centering 
\epsfxsize=6cm 
\epsfbox{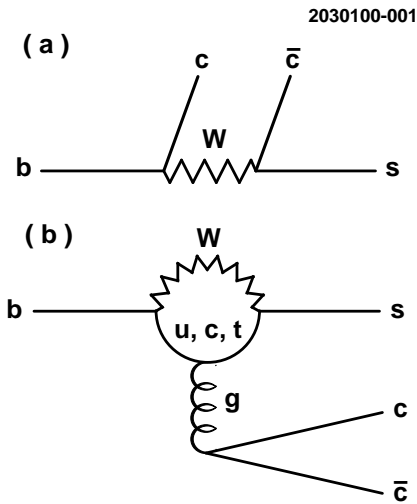} 
\caption{Tree (a) and penguin (b) diagrams for the  $b \to c \bar c s$
transition.} 
\label{fig:feynman-diagrams}
\end{figure}

$CP$ violation arises naturally in the Standard Model with three quark
 generations~\cite{Kobayashi:1973};  however,  it still remains one of
 the least  experimentally constrained sectors  of the Standard Model.
 Decays of $B$   mesons promise  to be   a  fertile  ground  for  $CP$
 violation studies.  Direct $CP$ violation, also called $CP$ violation
 in  decay,   occurs  when  the   amplitude  for   a   decay  and  its
 $CP$-conjugate   process   have different   magnitudes.   Direct $CP$
 violation  can be  observed in  both  charged  and neutral $B$  meson
 decays. At least  two interfering amplitudes with  different $CP$-odd
 (weak) and  $CP$-even (strong  or   electromagnetic) phases  are  the
 necessary ingredients   for  direct $CP$ violation.  For   the decays
 governed by the $b \to c \bar c s$ quark transition, such as $B^{\pm}
 \to J/\psi \,  K^{\pm}$  and  $B^0 (\overline  {B}^0)  \to  J/\psi \,
 K^0_S$,  there are  interfering    Standard Model  tree   and penguin
 amplitudes (Fig.~\ref{fig:feynman-diagrams}).  These amplitudes could
 have a significant relative strong  phase.  The relative weak  phase,
 however, is expected  to be  very small~\cite{Nir:1999}.   Therefore,
 the $CP$ asymmetry in $B^{\pm} \to J/\psi \, K^{\pm}$ decay is firmly
 predicted  in  the Standard  Model to  be  much smaller  than the 4\%
 precision of our measurement. 

A  $CP$  asymmetry  of ${\cal O}(10\%)$   in  $B^{\pm} \to   J/\psi \,
K^{\pm}$   decay is  possible in  a   specific two-Higgs doublet model
described in Ref.\cite{Soni}; such a large asymmetry could be measured
with our current data.  In  order to constrain any  of the New Physics
models, however, we need to know  the relative strong phases which are
difficult to determine.

The measurement   of  the $CP$ asymmetry  in  $B^0(\overline {B}^0)\to
J/\psi \,   K^0_S$ decay allows an  extraction  of the  relative phase
between the $B^0 - \overline {B}^0$ mixing amplitude and  the $b \to c
\bar c s$  decay  amplitude~\cite{sin2b}. In  the Standard Model  this
phase is equal  to $\sin 2\beta$, where $\beta  \equiv {\rm Arg} \left
(-  V_{cd}V^*_{cb}/V_{td}V^*_{tb}  \right)$.   An observation of  $CP$
asymmetry in  $B^{\pm} \to J/\psi \, K^{\pm}$  decay at a few per cent
or larger level will be a clear evidence for sources of $CP$ violation
beyond the Standard Model.  Such an  observation will also mean that a
measurement of  the $CP$ asymmetry  in $B^0(\overline {B}^0)\to J/\psi
\, K^0_S$ decay no longer determines $\sin 2\beta$. 

If some  mechanism causes direct $CP$ violation  to  occur in $B^{\pm}
\to J/\psi \, K^{\pm}$ decays, then the  same mechanism could generate
a $CP$  asymmetry  in $B^{\pm} \to \psi(2S)   \, K^{\pm}$ mode.  Final
state strong interactions,   however, could  be quite  different   for
$J/\psi  \, K$ and     $\psi(2S) \, K$    states; thus,  we   measured
$CP$-violating charge  asymmetries separately for $B^{\pm}  \to J/\psi
\, K^{\pm}$ and $B^{\pm} \to \psi(2S) \, K^{\pm}$ decay modes. 

The  data used   for our measurement   were collected  at the  Cornell
Electron Storage   Ring (CESR) with   two  configurations of the  CLEO
detector          called          CLEO~II~\cite{Kubota:1992ww}     and
CLEO~II.V~\cite{Hill:1998ea}.   The components  of the  CLEO  detector
most relevant   to this analysis   are  the charged particle  tracking
system, the CsI   electromagnetic calorimeter, and  the muon chambers.
In CLEO~II the momenta of charged particles are measured in a tracking
system  consisting of a    6-layer  straw tube  chamber, a    10-layer
precision  drift  chamber,  and  a  51-layer  main drift  chamber, all
operating inside a  1.5 T  solenoidal  magnet. The  main drift chamber
also provides a measurement of the  specific ionization, $dE/dx$, used
for particle identification.  For   CLEO~II.V, the straw  tube chamber
was replaced  with a 3-layer silicon vertex  detector, and  the gas in
the main   drift  chamber  was  changed  from  an  argon-ethane  to  a
helium-propane mixture.   The muon  chambers consist   of proportional
counters placed at increasing depth in steel absorber. 

 For this measurement we used 9.2~$\rm fb^{-1}$ of $e^+e^-$ data taken
 at the $\Upsilon(4S)$  resonance  and 4.6~$\rm fb^{-1}$ taken  60~MeV
 below the   $\Upsilon(4S)$ resonance. In  $\Upsilon(4S)$ decays $B^+$
 mesons are born only in pairs with $B^-$  mesons, therefore $B^+$ and
 $B^-$ mesons are  produced in equal numbers. Two  thirds of  the data
 used were collected with the CLEO~II.V detector.  The simulated event
 samples    used    in    this analysis      were  generated    with a
 GEANT-based~\cite{GEANT} simulation of the CLEO detector response and
 were processed in a similar manner as the data.

We  reconstructed $\psi^{(\prime)} \to  e^+ e^-$  and $\psi^{(\prime)}
\to \mu^+ \mu^-$   decays, where $\psi^{(\prime)}$  stands  for either
$J/\psi$  or $\psi(2S)$.  We  also   reconstructed $\psi(2S)$ in   the
$\psi(2S) \to J/\psi \, \pi^+ \pi^-$ channel. 

  Electron candidates were identified based on the  ratio of the track
  momentum to the associated shower  energy in the CsI calorimeter and
  on the  specific ionization in the drift  chamber. We recovered some
  of  the bremsstrahlung photons by  selecting  the photon shower with
  the smallest opening  angle  with respect to  the direction  of  the
  $e^\pm$ track evaluated at the interaction point, and then requiring
  this opening angle to be smaller than  $5^\circ$. We therefore refer
  to the $e^+  (\gamma) e^- (\gamma)$  invariant mass when we describe
  the $\psi^{(\prime)} \to e^+ e^-$ reconstruction. 

 For the $\psi^{(\prime)} \to  \mu^+ \mu^-$ reconstruction, one of the
muon  candidates  was required to penetrate  the  steel absorber  to a
depth  greater than 3  nuclear interaction  lengths.   We  relaxed the
absorber penetration requirement  for the second  muon candidate if it
was not expected to reach a muon chamber either because its energy was
too low  or because  it  did not point  to a  region of   the detector
covered by the muon  chambers.  For these  muon candidates we required
the ionization signature in the CsI  calorimeter to be consistent with
that of a muon. 
 
We extensively   used normalized    variables,  taking advantage    of
well-understood  track  and    photon-shower four-momentum  covariance
matrices   to    calculate   the    expected    resolution for    each
combination. The use of normalized  variables allows uniform candidate
selection  criteria to  be  applied to   the data  collected with  the
CLEO~II and CLEO~II.V  detector configurations.  The $\psi^{(\prime)}$
candidates  were selected using the   normalized invariant mass.   For
example, the  normalized $\mu^+  \mu^-$ invariant  mass  is defined as
$[M(\mu^+           \mu^-)-M_{\psi^{(\prime)}}]/\sigma(M)$,      where
$M_{\psi^{(\prime)}}$  is the world average value   of the $J/\psi$ or
$\psi(2S)$ mass~\cite{PDG}  and  $\sigma(M)$ is  the calculated   mass
resolution for that particular $\mu^+  \mu^-$ combination. The average
$\ell^+   \ell^-$     invariant  mass  resolution     is approximately
12~MeV$/c^2$.  We required the  normalized  $\mu^+ \mu^-$ mass  to  be
from $-4$ to 3 for  $J/\psi\to \mu^+ \mu^- $  candidates and from $-3$
to  3 for  $\psi(2S) \to   \mu^+ \mu^-$  candidates.  We required  the
normalized $e^+ (\gamma) e^- (\gamma)$ mass to be from  $-10$ to 3 for
$J/\psi\to e^+  e^-$ candidates and from $-3$  to 3 for  $\psi(2S) \to
e^+ e^-$  candidates.   For each $\psi^{(\prime)} \to   \ell^+ \ell^-$
candidate, we  performed  a fit constraining   its mass  to  the world
average value.  We  selected the $\psi(2S)  \to J/\psi \, \pi^+ \pi^-$
candidates by requiring the absolute  value of the normalized  $J/\psi
\, \pi^+ \pi^-$ mass  to be less than  3  and by requiring the  $\pi^+
\pi^-$ invariant mass to be  greater than 400~MeV/$c^2$.  The  average
$J/\psi   \,  \pi^+     \pi^-$  mass resolution      is  approximately
3~MeV/$c^2$. For each  $\psi(2S)\to J/\psi \, \pi^+  \pi^-$ candidate,
we performed a  fit constraining its mass to  the world average value.
Well-measured   tracks consistent  with   originating at the  $e^+e^-$
interaction point were selected as the  $K^{\pm}$ candidates. In order
avoid  any  additional    charge-correlated systematic  bias    in the
$K^{\pm}$  selection, we did   not impose any  particle identification
requirements on the $K^{\pm}$ candidates. 
 
\begin{figure}[htbp]
\centering \epsfxsize=14cm \epsfbox{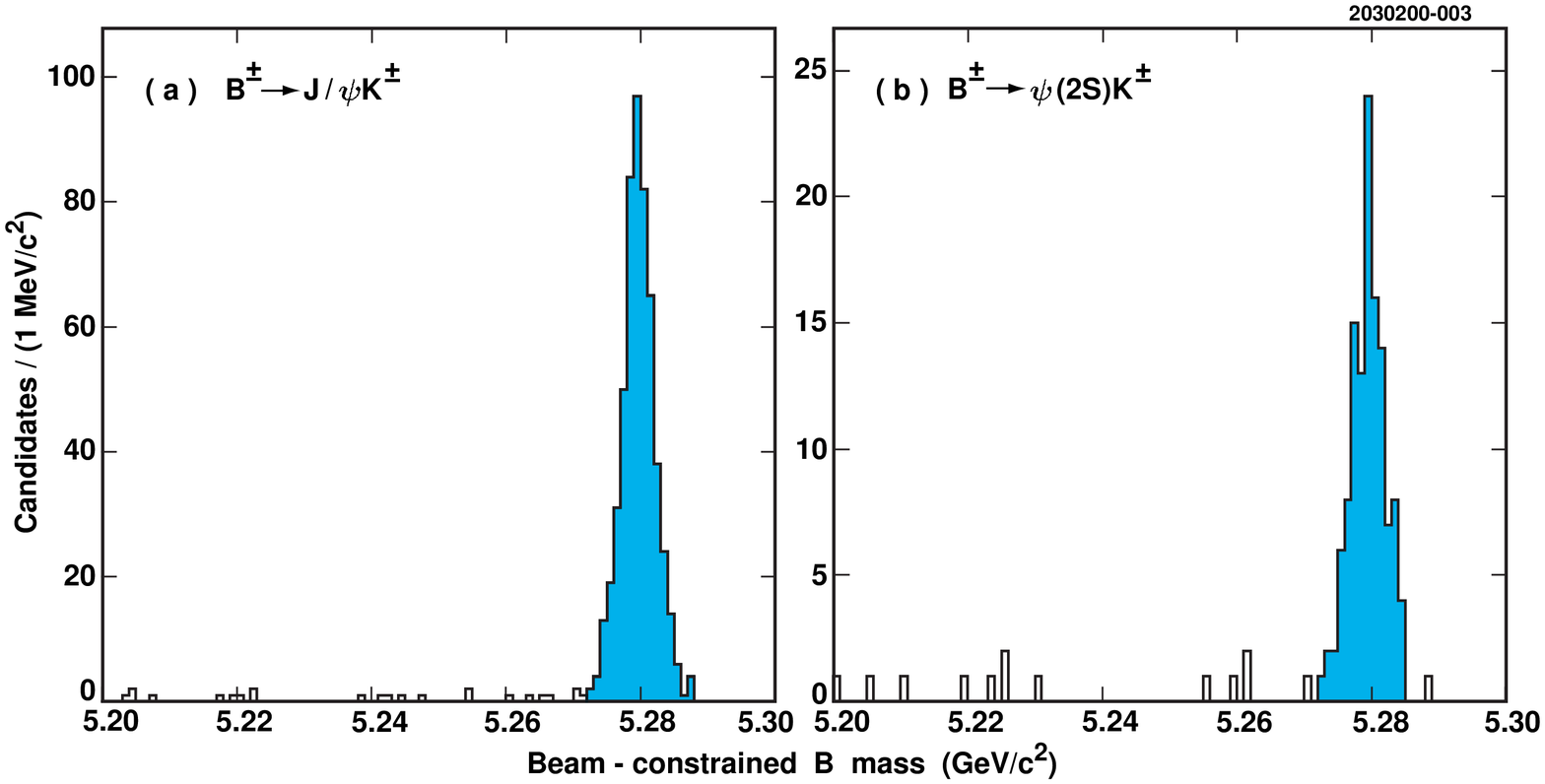} 
\caption{ Beam-constrained $B$  mass distribution for (a) $B^{\pm} \to
J/\psi  \,   K^{\pm}$  and  (b)  $B^{\pm}  \to  \psi(2S)   \, K^{\pm}$
candidates  passing   the       $|\Delta  E|/\sigma(\Delta       E)<3$
requirement.  The  shaded parts  of the  histograms represent  the 534
$B^{\pm} \to  J/\psi \, K^{\pm}$   and  120 $B^{\pm} \to  \psi(2S)  \,
K^{\pm}$    candidates    that   pass the     $|M(B)-M_B|/\sigma(M)<3$
requirement.} 
\label{fig:mb_all_data}
\end{figure}

 The    $B^{\pm} \to J/\psi \,   K^{\pm}$  and $B^{\pm} \to \psi(2S)\,
K^{\pm}$  candidates  were selected by   means of two observables. The
first observable is the difference between the energy of the $B^{\pm}$
candidate  and the beam  energy, $\Delta E  \equiv E(B^{\pm}) - E_{\rm
beam}$.  The  average resolution in $\Delta E$   is 10~MeV (8~MeV) for
the   $B^{\pm} \to  J/\psi \,   K^{\pm}$  ($B^{\pm}  \to \psi(2S)   \,
K^{\pm}$) candidates. We used the normalized $\Delta E$ observable for
candidate selection and required $|\Delta E|/\sigma(\Delta E)<3$.  The
second   observable   is    the      beam-constrained   $B$      mass,
$M(B)\equiv\sqrt{E^2_{\rm  beam}-p^2(B)}$,    where $p(B)$    is   the
magnitude of the $B$ candidate  momentum. The resolution in $M(B)$ for
the   $B^{\pm}     \to  \psi^{(\prime)} \, K^{\pm}$     candidates  is
2.7~MeV/$c^2$ and is dominated by  the beam energy spread.  The $M(B)$
distributions for the $B^{\pm} \to J/\psi \, K^{\pm}$ and $B^{\pm} \to
\psi(2S)\, K^{\pm}$  candidates  passing the $|\Delta E|/\sigma(\Delta
E)<3$ requirement  are shown  in Fig.~\ref{fig:mb_all_data}.  We  used
the normalized $M(B)$  observable for candidate selection and required
$|M(B)-M_B|/\sigma(M)<3$. 
 
The $CP$-violating charge asymmetry in $B^{\pm} \to J/\psi \, K^{\pm}$
decays is defined as a branching fraction asymmetry 
\begin{eqnarray*}
\nonumber {\cal A}_{CP} \equiv \frac{ {\cal B}(B^- \to J/\psi \, K^-)-
{\cal  B}(B^+ \to J/\psi  \, K^+)} {{\cal B}(B^-  \to  J/\psi \, K^-)+
{\cal B}(B^+ \to J/\psi \, K^+)} \;. 
\end{eqnarray*}
In     this    definition we    adopted  the    sign   convention from
Ref.~\cite{Acp-CLEO}.  The  same definition is   used for $B^{\pm} \to
\psi(2S) \, K^{\pm}$ mode. 
  
Table~I lists signal yields together with observed charge asymmetries.
The possible sources of systematic  uncertainty and bias in the ${\cal
A}_{CP}$ measurement are described below. 
\begin{table}[htbp]
\label{tab:counting}
\caption{Number of selected candidates, the observed charge asymmetry,
and the corrected asymmetry.} 
\begin{center}
\begin{tabular}{l c c c c c } 
 Mode & $N(B^{\pm})$ & $N(B^{-})$ & $N(B^{+})$ & {\large
$\frac{N(B^{-})-N(B^{+})}{N(B^{-})+N(B^{+})}$} & $ {\cal A}_{CP}$ \\
\hline $B^{\pm} \to J/\psi \, K^{\pm}$ & 534 & 271 & 263 & $(+1.5\pm
4.3)\%$ & $(+1.8\pm4.3[{\rm stat}]\pm 0.4[{\rm syst}])\%$ \\ $B^{\pm}
\to \psi(2S) \, K^{\pm}$ & 120 & 61 & 59 & $(+1.7\pm 9.1)\%$ &
$(+2.0\pm9.1[{\rm stat}]\pm 1.0[{\rm syst}])\%$ \\ 
\end{tabular}
\end{center}
\end{table}

{\it Background.}     --- From  fits to    the   beam-constrained mass
distributions    (Fig.~\ref{fig:mb_all_data}),   we  estimated     the
combinatorial     background            to     be  $3.5^{+2.8}_{-1.7}$
($1.7^{+2.0}_{-1.0}$)  for $B^{\pm} \to   J/\psi \, K^{\pm}$ ($B^{\pm}
\to   \psi(2S) \,  K^{\pm}$) mode.  The   background from $B^{\pm} \to
\psi^{(\prime)} \, \pi^{\pm}$ decays has to  be added because $B^{\pm}
\to  \psi^{(\prime)}  \,  \pi^{\pm}$   candidates  contribute  to  the
beam-constrained mass peaks.  Using simulated events, we estimated the
background from  $B^{\pm} \to \psi^{(\prime)}  \, \pi^{\pm}$ decays to
be  $1.5\pm0.5$ events  for $B^{\pm}  \to J/\psi  \,  K^{\pm}$ and 0.1
event for  $B^{\pm}  \to \psi(2S) \,  K^{\pm}$ mode.    We assumed the
branching ratio of ${\cal   B}(B^{\pm} \to J/\psi  \, \pi^{\pm})/{\cal
B}(B^{\pm} \to  J/\psi \, K^{\pm})=(5.1\pm1.4)\%$ \cite{PDG}; the same
value was assumed  for  $B^{\pm}  \to \psi(2S) \,  \pi^{\pm}$  decays.
Total background is therefore estimated to be $5^{+3}_{-2}$ events for
$B^{\pm} \to J/\psi \, K^{\pm}$ and  $2^{+2}_{-1}$ events for $B^{\pm}
\to \psi(2S) \,   K^{\pm}$  mode.  As a  check,   we used samples   of
simulated   events together with  the   data  collected below  the  $B
\overline B$ production threshold and estimated total background to be
$3.3\pm0.8$ events for $B^{\pm} \to J/\psi \, K^{\pm}$ and $3.7\pm0.9$
events for $B^{\pm}  \to \psi(2S) \, K^{\pm}$  mode.  We verified that
the simulation   accurately  reproduced the rate  and  distribution of
candidates in  the data in the $\Delta  E$ vs.  $M(B)$ plane near, but
not   including, the signal  region.    Backgrounds are expected to be
$CP$-symmetric. We measured the charge asymmetry for the candidates in
the side-band regions of the $\Delta E$ and $M(B)$ distributions to be
$(+2.2\pm4.1)\%$   for $B^{\pm}   \to      J/\psi  \, K^{\pm}$     and
$(-1.2\pm6.4)\%$ for  $B^{\pm}   \to \psi(2S) \,  K^{\pm}$.    We also
verified  that our  final result  does  not  critically  depend on the
assumption of zero $CP$  asymmetry for background events.  We  assumed
that the number  of background events  entering  our sample follows  a
Poisson distribution with  a mean of 5 events  for $B^{\pm} \to J/\psi
\, K^{\pm}$  and  4  events  for  $B^{\pm}  \to  \psi(2S)  \, K^{\pm}$
mode. We also assumed that the $CP$-violating charge asymmetry for the
background is $+30\%$. Using   Monte Carlo techniques, we  found  that
background with such  properties introduces a $+0.3\%$ ($+1.0\%$) bias
in our ${\cal  A}_{CP}$  measurement for the  $B^{\pm}  \to J/\psi  \,
K^{\pm}$ ($B^{\pm} \to \psi(2S)  \,  K^{\pm}$)  mode.  We assigned   a
systematic uncertainty on ${\cal A}_{CP}$ of  $0.3\%$ for $B^{\pm} \to
J/\psi \, K^{\pm}$ and $1.0\%$ for $B^{\pm} \to \psi(2S) \, K^{\pm}$. 

{\it Charge asymmetry   for   inclusive tracks.} ---  Collisions    of
 particles with  the nuclei  in   the detector material   occasionally
 result in recoil protons, but almost never in recoil antiprotons.  To
 fake a $K^+$ candidate, a recoil proton has  to have a momentum of at
 least 1.2~GeV/$c$ and its track should be consistent with originating
 at the $e^+e^-$ interaction point.   In order to  study the effect of
 possible recoil proton contamination of our $K^+$ sample, we selected
 inclusive tracks satisfying  the same track  quality criteria  as for
 the charged  kaon candidates in  the  $B^{\pm} \to \psi^{(\prime)} \,
 K^{\pm}$  reconstruction.  The kaon momentum  in the laboratory frame
 is  between 1.2 and  1.4~GeV/$c$  for  the $B^{\pm}  \to \psi(2S)  \,
 K^{\pm}$ mode and between 1.55 and 1.85 GeV/$c$  for the $B^{\pm} \to
 J/\psi  \, K^{\pm}$  mode. We  have indeed  found more  positive than
 negative tracks  in these two momentum ranges.    For all tracks with
 momentum  between  1.2 and 1.4   GeV/$c$, we  have observed a  charge
 asymmetry    of  $(N^-  -  N^+)/(N^-   +  N^+)=(-0.22\pm0.03)\%$; the
 corresponding number   for tracks  with   momentum  between 1.55  and
 1.85~GeV/$c$ is $(-0.17\pm0.04)\%$. Besides increasing our confidence
 that   our    track  reconstruction  procedure   does  not  introduce
 significant charge-correlated bias, this study also confirms that the
 number of  recoil  protons entering the  pool of  $K^+$ candidates is
 negligible even before  the reconstruction of  the full $B^{\pm}  \to
 \psi^{(\prime)}   \,  K^{\pm}$ decay chain.   We   did not assign any
 systematic uncertainty. 

{\it Difference  in $K^+$ vs.  $K^-$ detection efficiencies.}  --- The
flavor of the  $B$ meson is tagged by  the charged kaon; therefore, we
searched for  charge-correlated  systematic  bias associated with  the
$K^{\pm}$ detection and momentum measurement.  The cross sections  for
nuclear  interactions are larger for  negative than for positive kaons
from  $B^{\pm} \to \psi^{(\prime)} \,   K^{\pm}$ decays.  We used  two
methods to  evaluate  the difference  in  $K^+$  vs.  $K^-$  detection
efficiencies. In the first method we performed an analytic calculation
of the   expected  asymmetry, combining  the   data   on the   nuclear
interaction cross  sections for the  $K^+$ and $K^-$ mesons~\cite{PDG}
with the  known composition  of  the CLEO detector material.   In  the
second method we used the GEANT-based simulation  of the CLEO detector
response, processing  the simulated events in  a similar manner as the
data.  Both   methods  are in   excellent   agreement  that  the $K^+$
reconstruction efficiency  is approximately  $0.6\%$  higher  than the
$K^-$ reconstruction efficiency.  The corresponding  charge-correlated
detection efficiency asymmetry   is therefore $-0.3\%$.   We applied a
$+0.3\%$ correction to the measured values of ${\cal A}_{CP}$ both for
$B^{\pm} \to   J/\psi \, K^{\pm}$ and  for  $B^{\pm} \to   \psi(2S) \,
K^{\pm}$  modes. We assigned 100\%  of the correction  as a systematic
uncertainty. 

{\it Bias in $K^+$ vs. $K^-$ momentum measurement.} --- This bias will
separate the $\Delta E \equiv E(B^{\pm})-E_{\rm beam}$ peaks for $B^+$
and   $B^-$ candidates  so  that  the requirement   on $\Delta  E$ can
manifest a  preference for the  $B$ candidates of  a certain  sign. We
measured the difference  in mean $\Delta E$  for  the $B^+$ and  $B^-$
candidates to be $0.6\pm0.8$~MeV.  This result is consistent with zero
and very  small compared to the  approximately $\pm30$~MeV window used
in the $\Delta E$ requirement.  We also used high-momentum muon tracks
from $e^+ e^- \to \mu^+ \mu^-$ events as  well as samples of $D^0$ and
$D^{\pm}_{(s)}$ meson  decays~\cite{Acp-CLEO}  to put stringent limits
on possible charge-correlated  bias in  the momentum measurement.   We
conclude that the bias in $K^+$ vs.   $K^-$ momentum reconstruction is
negligible for our $CP$-violation measurement. 

In conclusion, we have measured the $CP$-violating charge asymmetry to
be $(+1.8\pm4.3[ {\rm stat}]\pm 0.4[{\rm  syst}])\%$ for $B^{\pm}  \to
J/\psi   \,  K^{\pm}$   and   $(+2.0\pm9.1[ {\rm   stat}]\pm  1.0[{\rm
syst}])\%$  for $B^{\pm} \to \psi(2S)  \,  K^{\pm}$.  These values  of
${\cal A}_{CP}$ include a $+0.3\%$ correction due to a slightly higher
reconstruction  efficiency for the positive   kaons.  Our results  are
consistent with the Standard  Model expectations and provide the first
experimental   test of the assumption  that   direct $CP$ violation is
negligible in $B \to \psi^{(\prime)} \, K$ decays. 

We gratefully acknowledge the effort of the CESR staff in providing us
with excellent luminosity  and  running conditions.  We  thank A.~Soni
and M.~Neubert for useful discussions.  This work was supported by the
National  Science Foundation,  the  U.S.   Department of  Energy,  the
Research Corporation,  the Natural  Sciences and  Engineering Research
Council   of Canada, the   A.P.  Sloan Foundation,  the Swiss National
Science Foundation, and the Alexander von Humboldt Stiftung.

\end{document}